\documentstyle[emulateapj,apjfonts,psfig,graphicx,cases]{article}

\def\grb{GRB\,020124}
\def\ra#1#2#3{#1$^{\rm h}$#2$^{\rm m}$#3$^{\rm s}$}
\def\dec#1#2#3{$#1^\circ#2'#3''$}

\begin{document}

\def\cit{1}
\def\mso{2}
\def\vla{3}
\def\uva{4}
\def\jhu{5}
\def\rom{6}
\def\cnr{7}
\def\fer{8}
\def\col{9}
\def\nms{10}
\def\ucb{11}
\def\god{12}
\def\car{13}
\def\mit{14}
\def\tap{15}
\def\uta{16}

\title{\large The Faint Optical Afterglow and Host Galaxy of \grb:
Implications for the Nature of Dark Gamma-Ray Bursts}

\author{
E.    Berger\altaffilmark{\cit},
S. R. Kulkarni\altaffilmark{\cit},
J. S. Bloom\altaffilmark{\cit},
P. A. Price\altaffilmark{\cit,\mso},
D. W. Fox\altaffilmark{\cit},
D. A. Frail\altaffilmark{\cit,\vla},
T. S. Axelrod\altaffilmark{\mso},
R. A. Chevalier\altaffilmark{\uva},
E.    Colbert\altaffilmark{\jhu},
E.    Costa\altaffilmark{\rom},
S. G. Djorgovski\altaffilmark{\cit},
F.    Frontera\altaffilmark{\cnr,\fer},
T. J. Galama\altaffilmark{\cit},
J. P. Halpern\altaffilmark{\col},
F. A. Harrison\altaffilmark{\cit},
J.    Holtzman\altaffilmark{\nms},
K.    Hurley\altaffilmark{\ucb},
R. A. Kimble\altaffilmark{\god},
P. J. McCarthy\altaffilmark{\car},
L.    Piro\altaffilmark{\rom},
D.    Reichart\altaffilmark{\cit},
G. R. Ricker\altaffilmark{\mit}, 
R.    Sari\altaffilmark{\tap},
B. P. Schmidt\altaffilmark{\mso},
J. C. Wheeler\altaffilmark{\uta},
R.    Vanderppek\altaffilmark{\mit}
\&\ S. A. Yost\altaffilmark{\cit}
}

\altaffiltext{\cit}{Division of Physics, Mathematics and Astronomy,
        105-24, California Institute of Technology, Pasadena, CA
        91125} 
\altaffiltext{\mso}{Research School of Astronomy \& Astrophysics, 
        Mount Stromlo Observatory, via Cotter Rd., Weston Creek 2611,
        Australia} 
\altaffiltext{\vla}{National Radio Astronomy Observatory, Socorro,
        NM 87801} 
\altaffiltext{\uva}{Department of Astronomy, University of Virginia,
        P.O. Box 3818, Charlottesville, VA 22903-0818}
\altaffiltext{\jhu}{Department of Physics and Astronomy, Johns Hopkins 
        University, 3400 N.~Charles St., Baltimore, MD  21218}
\altaffiltext{\rom}{Istituto Astrofisica Spaziale, C.N.R., Area di
        Tor Vergata, Via Fosso del Cavaliere 100, 00133 Roma, Italy}
\altaffiltext{\cnr}{Istituto Astrofisica Spaziale and Fisica Cosmica,
        C.N.R., Via Gobetti, 101, 40129 Bologna, Italy}
\altaffiltext{\fer}{Physics Department, University of Ferrara, Via
        Paradiso, 12, 44100 Ferrara, Italy}
\altaffiltext{\col}{Columbia Astrophysics Laboratory, Columbia
        University, 550 West 120th Street, New York, NY 10027}
\altaffiltext{\nms}{Department of Astronomy, MSC 4500, New Mexico
        State University, P.O.~Box 30001, Las Cruces, NM 88003}
\altaffiltext{\ucb}{University of California at Berkeley, Space
        Sciences Laboratory, Berkeley, CA 94720-7450}
\altaffiltext{\god}{Laboratory for Astronomy and Solar Physics, NASA
        Goddard Space Flight Center, Code 681, Greenbelt, MD 20771}
\altaffiltext{\car}{Carnegie Observatories, 813 Santa Barbara Street,
        Pasadena, CA 91101}
\altaffiltext{\mit}{Center for Space Research, Massachusetts Institute
        of Technology, 70 Vassar Street, Cambridge, MA 02139-4307}
\altaffiltext{\tap}{Theoretical Astrophysics 130-33, California
        Institute of Technology, Pasadena, CA 91125}
\altaffiltext{\uta}{Astronomy Department, University of Texas,
        Austin, TX 78712}

\begin{abstract}

We present ground-based optical observations of \grb{} starting 1.6
hours after the burst, as well as subsequent Very Large Array (VLA)
and {\it Hubble Space Telescope} (HST) observations.  The optical
afterglow of \grb{} is one of the faintest afterglows detected to
date, and it exhibits a relatively rapid decay, $F_\nu\propto
t^{-1.60\pm 0.04}$, followed by further steepening.  In addition, a
weak radio source was found coincident with the optical afterglow.
The HST observations reveal that a positionally coincident host galaxy
must be the faintest host to date, $R\gtrsim 29.5$ mag.  The afterglow
observations can be explained by several models requiring little or no
extinction within the host galaxy, $A_V^{\rm host}\approx 0-0.9$ mag.   
These observations have significant implications for the statistics of
the so-called dark bursts (bursts for which no optical afterglow is
detected), which are usually attributed to dust extinction within the
host galaxy.  The faintness and relatively rapid decay of the
afterglow of \grb{}, combined with the low inferred extinction
indicate that some dark bursts are intrinsically dim and not dust
obscured.  Thus, the diversity in the underlying properties of optical
afterglows must be observationally determined before substantive
inferences can be drawn from the statistics of dark bursts.
\end{abstract}


\section{Introduction}
\label{sec:intro}

One of the main observational results stemming from five years of
$\gamma$-ray burst (GRB) follow-ups at optical wavelengths is that
about $60\%$ of well-localized GRBs lack a detected optical afterglow,
(``dark bursts''; Taylor et al. 2000; Fynbo et al. 2001; Reichart \&
Yost 2001; Lazzati, Covino, \& Ghisellini
2002)\nocite{tbf+00,fjg+01,ry01,lcg02}.  In some cases, a
non-detection of the optical afterglow could simply be due to a
failure to image quickly and/or deeply enough.  However, there are two
GRBs for which there is strong evidence that the optical emission
should have been detected, based on an extrapolation of the radio and
X-ray emission (Djorgovski et al. 2001a; Piro et
al. 2002)\nocite{dfk+01a,pfg+02}.  One interpretation in these two
cases is that the optical light was extinguished by dust, either
within the immediate environment of the burst or elsewhere along the
line of sight (e.g.~Groot et al. 1998)\nocite{ggv+98}.  An alternative
explanation is a high redshift, leading to absorption of the optical
light in the Ly$\alpha$ forest.  However, the redshifts of the
underlying host galaxies of these GRBs are of order unity (Djorgovski
et al. 2001a; Piro et al. 2002)\nocite{dfk+01a,pfg+02}.

Several authors have recently argued that a large fraction of the dark
bursts are due to dust extinction within the local environment of the
bursts (e.g.~Reichart \& Yost 2001; Lazzati et al. 2002; Reichart \&
Price 2002)\nocite{ry01,lcg02,rp02}, but other scenarios have also
been suggested (e.g.~Lazzati et al. 2002).  Moreover, it has been
noted that regardless of the location of extinction within the host
galaxy, the fraction of dark bursts is a useful upper limit on the
fraction of obscured star formation (Kulkarni et al. 2000;
Djorgovski et al. 2001b; Ramirez-Ruiz, Trentham, \& Blain 2002;
Reichart \& Price 2002)\nocite{kbb+00,dkb+01,rtb02,rp02}.

However, from an observational point of view, we must have a clear
understanding of the diversity of afterglow properties before
extracting astrophysically interesting inferences from dark bursts.
For example, afterglows which are faint or fade rapidly (relative to
the detected population) would certainly bias the determination of the
fraction of truly obscured bursts.  In this vein, Fynbo et al. (2001),
noting the faint optical afterglow of GRB\,000630, argued that some
dark bursts are due to a failure to image deeply and/or quickly
enough, rather than dust extinction.

Here we present optical and radio observations of \grb{}, an afterglow
that would have been classified dark had it not been
for rapid and deep searches.  Furthermore, \grb{} is an example of an
afterglow, which is dim due to the combination of intrinsic faintness
and a relatively fast decline, and not strong extinction.

\section{Observations}
\label{sec:obs}

\subsection{Ground-Based Observations}
\label{sec:opt}

\grb{}, localized by the HETE-II satellite on 2002, Jan 24.44531 UT,
had a duration of $\sim 70$ s and a fluence ($6-400$ keV) of
$3\times 10^{-6}$ erg cm$^{-2}$ (Ricker et al. 2002)\nocite{rak+02}.  
Eight minutes after receiving the
coordinates\footnotemark\footnotetext{This corresponds to 1.6 hours
after the burst detection.} we observed the error box with the
dual-band ($B_M$, $R_M$) MACHO imager mounted on the robotic 50-in
telescope at the Mount Stromlo Observatory (MSO).  We also observed
the error box with the Wide-Field Imager on the 40-in telescope at
Siding Spring Observatory (SSO).  We were unable to identify a
transient source within the large error box (Price, Schmidt, \&
Axelrod 2002)\nocite{psa02}.

\begin{figure} 
\plotone{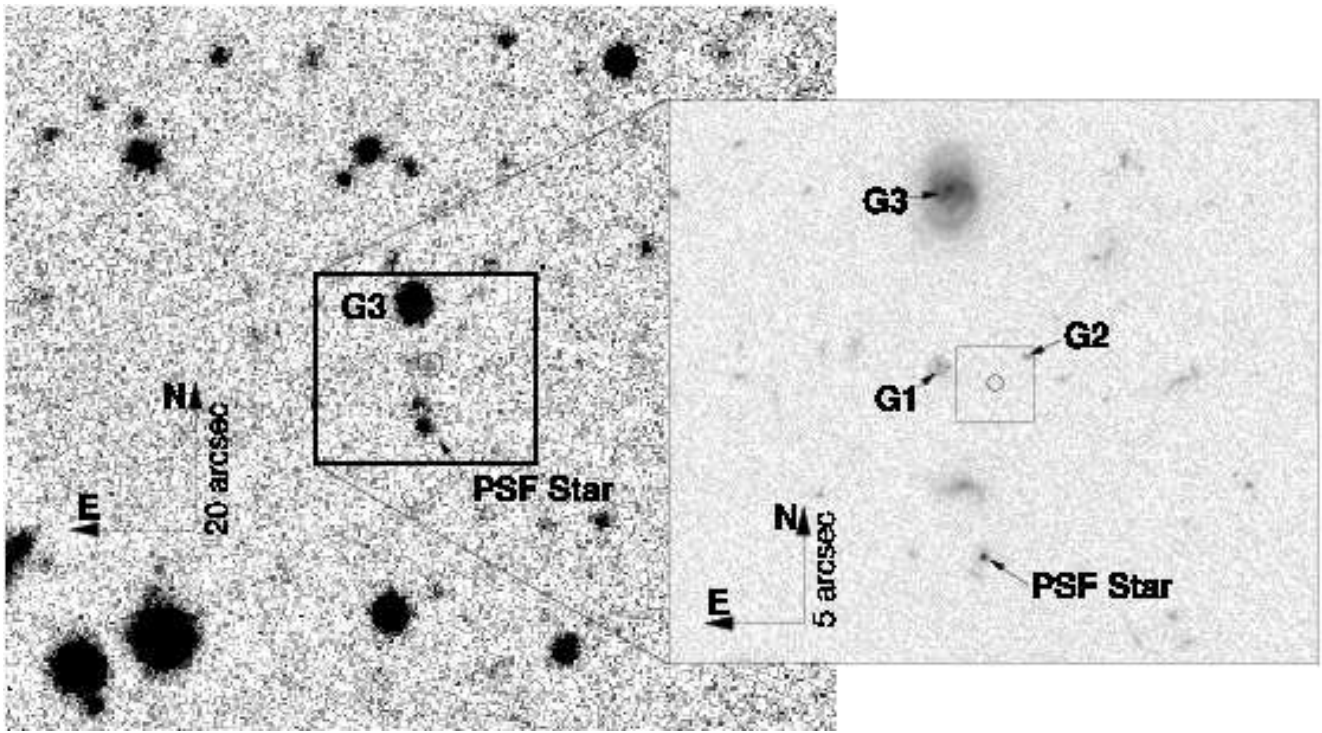}
\caption{Palomar 200-inch (left) and HST epoch 1 (inset) images of the
field of \grb{}. The OT is circled in both images. The OT was of
comparable brightness to G1 at the epoch of the P\,200 image and
significantly fainter than G1 three weeks later. The box overlaying
the inset shows the portion of the HST images depicted in
Figure~\ref{fig:hst}. Relevant sources described in the text are
noted. The HST image is shown with logarithmic scaling to highlight
the features of nearby galaxies..
\label{fig:field}}
\end{figure}

We subsequently observed the error box refined by the Inter-Planetary
Network (Hurley et al. 2002)\nocite{hcr+02} with the Palomar 48-in
Oschin Schmidt using the unfiltered NEAT imager.  PSF-matched image
subtraction (Alard 2000)\nocite{ala00} between the MACHO and NEAT
images revealed a fading source (Price et al. 2002)\nocite{pfy+02},
which was $R\approx 18$ mag at the epoch of our first observations,
and not present in the Digitized Sky Survey.  Two nights later we
observed the afterglow using the Jacobs CAMera (JCAM; Bloom et
al. 2002b)\nocite{bkc+02} mounted at the East arm focus of 
the Palomar 200-in telescope (Bloom et al. 2002a)\nocite{blo02}.  The
position of the fading source is $\alpha$(J2000)=\ra{9}{32}{50.78},
$\delta$(J2000)=\dec{-11}{31}{10.6}, with an uncertainty of about
$0.4$ arcsec in each coordinate (Fig.~\ref{fig:field}).

Using the Very Large Array (VLA\footnotemark\footnotetext{The VLA is
operated by the National Radio Astronomy Observatory, a facility of
the National Science Foundation operated under cooperative agreement
by Associated Universities, Inc.}) we observed the fading source at 
8.46 and 22.5 GHz (see Table~\ref{tab:rad}).  We detect a faint 
source, possibly fading, at 8.46 GHz located at
$\alpha$(J2000)=\ra{9}{32}{50.81}, 
$\delta$(J2000)= \dec{-11}{31}{10.6}, with an uncertainty of about
$0.1$ arcsec in each coordinate.  Given the positional coincidence
between the fading optical source and radio detection we suggest this
source to be the afterglow of \grb.

The optical images were bias-subtracted and flat-fielded in the
standard manner.  To extract the photometry we weighted the aperture
with a Gaussian equivalent to the seeing disk ("weighted-aperture
photometry"), using IRAF/{\tt wphot}.  The photometric zero-points
were set through photometry of calibrated field stars (Henden
2002)\nocite{hen02} with magnitudes transformed to the appropriate
system (Bessell \& Germany 1999; Smith et
al. 2002)\nocite{bg99,stk+02}.  The photometry is summarized in
Tab.~\ref{tab:opt}

\subsection{{\it Hubble Space Telescope} Observations}
\label{sec:hst}

We observed the afterglow with the {\it Hubble Space Telescope}
(HST) using the Space Telescope Imaging Spectrograph (STIS) on 2002
Feb.~11.09, 18.30, and 25.71 UT (Bloom et al. 2002a)\nocite{bpf+02},
as part of our large HST Cycle 10 program (GO-9180, PI: Kulkarni).
The HST observations consisted of 750--850 sec exposures.  The HST
data were retrieved
after ``On-The-Fly'' pre-processing.  Using IRAF we drizzled (Fruchter
\& Hook 2002)\nocite{fh02} each image onto a grid with pixels smaller
than the original by a factor of two and using {\tt pixfrac} of 0.7.  

We found an astrometric tie between the HST and JCAM images using
IRAF/{\tt geomap} with nine suitable astrometric tie objects in common
between the images.  The rms of the resultant mapping is 133 mas (RA)
and 124 mas (Dec).  Using this mapping and IRAF/{\tt geoxytrans} we
transfered the afterglow position on the JCAM image to the HST images.
The rms of the transformation is 604 mas (RA) and 596 mas (Dec), and is
dominated by the uncertainty in the JCAM position.

\begin{figure} 
\plotone{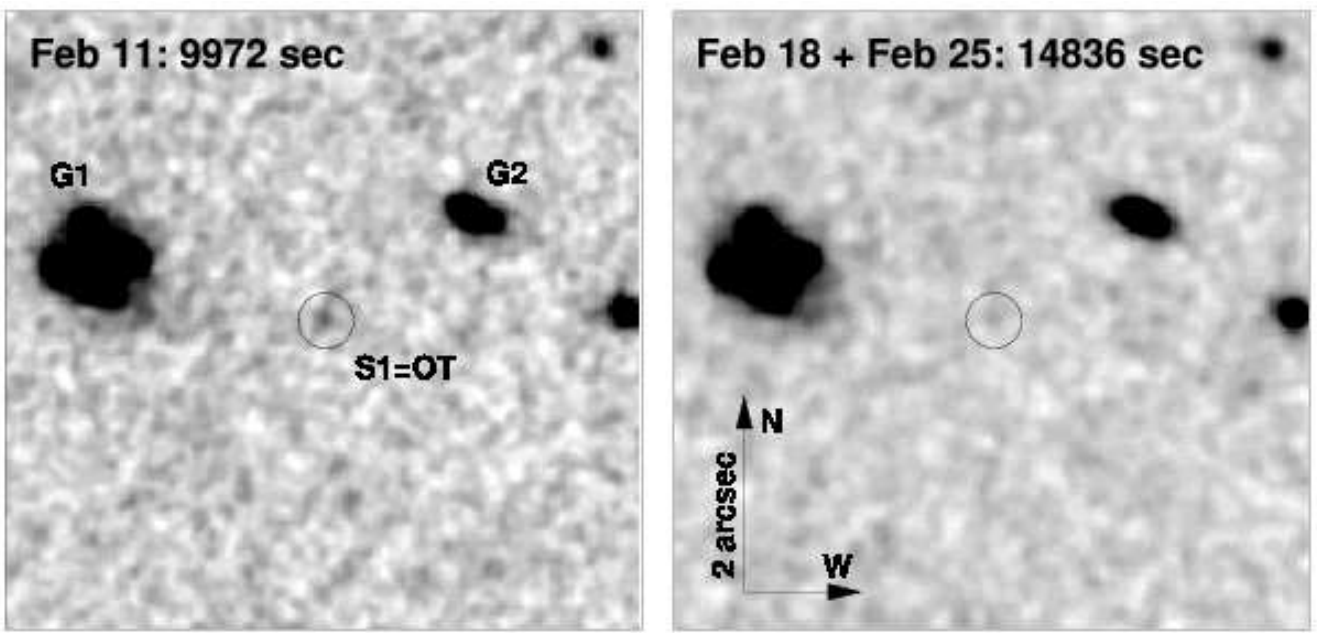}
\caption{The faint optical transient (OT) of \grb{} as viewed using
HST/STIS.  Shown are the summed, smoothed images from epoch 1 (left) and 
epochs 2+3 (right).  The greyscales have been matched such that a given
flux is represented by the same shade in each image.  The circle
is centered at the same sky position in both imagess.  Clearly, the source S1, 
identified with the position of the afterglow of \grb{} has faded.
\label{fig:hst}}
\end{figure}

The source "S1" (Fig.~\ref{fig:hst}) coincides with the afterglow
position within the astrometric uncertainty.  We performed
differential photometry at the position of S1 by registering the images
of epochs 1 and 2 using a cross-correlation of a field of size 10
arcsec centered on S1 (using IRAF/{\tt crosscor} and {\tt
shiftfind}).  We used IRAF/{\tt center} and the FWHM of a relatively
bright point source ("PSF star"; Fig.~\ref{fig:field}) to fix the
position of S1 in each of the final images, and to determine the
uncertainty in the position.

We photometered the source (and the PSF star) in epoch 1 using
IRAF/{\tt phot}, and determined a count-rate of $0.0921\pm 0.013$
e$^-$ s$^{-1}$ (corrected by $17\%$ for the loss of flux from an
infinite aperture radius).  Using IRAF/{\tt synphot} and assuming a
source spectrum of $f_\lambda \propto \lambda^{-1.4}$ (see below), we
find that the source was $R=28.55_{-0.14}^{+0.16}$ mag at the time of
epoch 1.  The photometry of the three epochs is summarized in
Tab.~\ref{tab:opt2}.  Please note that this more careful analysis
supersedes our preliminary report (Bloom et al. 2002c)\nocite{bpf+02}.

There are no obvious persistent sources within 1.75 arcsec of
the OT down to $R\approx 29.5$ mag.  To date, all of the GRBs
localized to sub-arcsecond accuracy have viable hosts brighter than
this level within $\sim 1.3$ arcsec of the OT position (Bloom,
Kulkarni, \& Djorgovski 2002)\nocite{bkd02}.  The faintest host to
date is that of GRB\,990510, $R\sim 28.5$ mag (z=1.619; Vreeswijk et
al. 2001)\nocite{vfk+01}.  Thus, the host of \grb{} may be at a
somewhat higher redshift; however, $z\lesssim 4.5$ since the afterglow
was detected in the $B_M$ filter.

\section{Modeling of the Optical Data}
\label{sec:model}

\begin{figure} 
\plotone{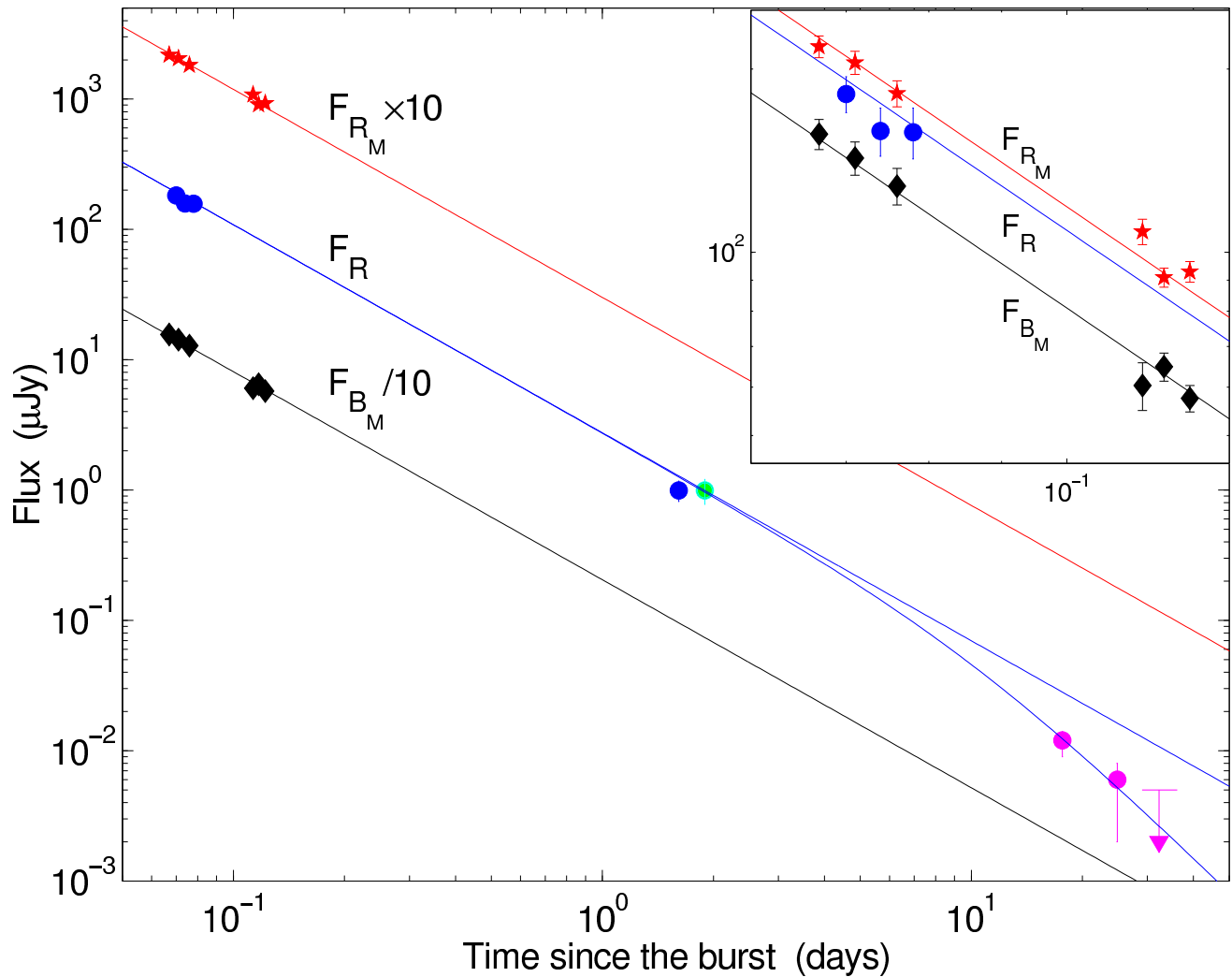}
\caption{Optical lightcurves of \grb{} (top to bottom: $R_M$, $R$, and
$B_M$), corrected for Galactic extinction, $E(B-V)=0.052$ mag
(Schlegel et al. 1998).  The solid lines are a representative jet
model (${\rm ISM/Wind_R}$; see Tab.~\ref{tab:models}), while the
dashed line is an extrapolation of the early evolution without a
break.  With no break in the $R$-band lightcurve, the predicted
magnitude at the epoch of the first HST observation exceeds the
measured values by $5\sigma$.  The flux measured in the last HST epoch
is plotted as a $2\sigma$ upper limit.
\label{fig:opt}}
\end{figure}

In Figure~\ref{fig:opt} we plot the optical lightcurves of \grb{},
including a correction for Galactic extinction, $E(B-V)=0.052$ mag
(Schlegel et al. 1998)\nocite{s+98}.  The optical lightcurves are 
usually modeled as $F_\nu(t,\nu) = F_{\nu,0} (t/t_0)^{\alpha}
(\nu/\nu_0)^{\beta}$.  However, as can be seen in Fig.~\ref{fig:opt},
the $R$-band lightcurve cannot be described by a single power law.
Restricting the fit to $t<2$ days we obtain ($\chi^2_{\rm min}=15$ for
14 degrees of freedom) $\alpha_1=-1.60\pm 0.04$, $\beta=-1.43\pm
0.14$, and $F_{\nu,0}=2.96\pm 0.25$ $\mu$Jy; here $F_{\nu,0}$ is
defined at the effective frequency of the $R_M$ filter and $t=1$ day.
For $t>2$ days we get $\alpha_2\approx -1.9$.

To account for the steepening we modify the model for the $R$-band
lightcurve to: 
\begin{equation}
F_\nu(t,\nu)=F_{\nu,0} (\nu/\nu_0)^{\beta} [(t/t_b)^{\alpha_1n}+
(t/t_b)^{\alpha_2n}]^{1/n},
\label{eqn:model2}
\end{equation}
where, $\alpha_1$ is the asymptotic index for $t\ll t_b$, $\alpha_2$
is the asymptotic index for $t\gg t_b$, $n<0$ provides a smooth
joining of the two asymptotic segments, and $t_b$ is the time at which
the asymptotic segments intersect.  We retain the simple model for the
$R_M$ and $B_M$ lightcurves since they are restricted to $t\lesssim
0.13$ days (i.e.~well before the observed steepening).

We investigate two alternatives for the observed steepening in the
framework of the afterglow synchrotron model (e.g.~Sari, Piran, \&
Narayan 1998).  In this framework, $\alpha_1$, $\alpha_2$, and $\beta$
are related to each other through the index ($p$) of the electron
energy distribution, $N(\gamma)\propto\gamma^{-p}$ (for $\gamma >
\gamma_{\rm min}$).  The relations for the models discussed below, as
well as the resulting closure relations, $\alpha_1+b\beta+c=0$, are
summarized in Tab.~\ref{tab:models}.

\subsection{Cooling Break}
\label{sec:cooling}

The observed steepening, $\Delta\alpha\equiv \alpha_2-\alpha_1\approx 
-0.3$, can be due to the passage of the synchrotron cooling
frequency, $\nu_c$, through the $R$-band.  This has been inferred, for
example, in the afterglow of GRB\,971214, at $t\sim 0.2$ days
(Wijers \& Galama 1999)\nocite{wg99}.  If the steepening is due to
$\nu_c$, this rules out models in which the ejecta expand into a
circumburst medium with $\rho\propto r^{-2}$ (hereafter, Wind),
because in this model $\nu_c$ increases with time ($\propto t^{1/2}$;
Chevalier \& Li 1999)\nocite{cl99}, and one expects $\Delta\alpha =
0.25$.  

There are two remaining models to consider in this case: (i) spherical
expansion into a circumburst medium with constant density (hereafter,
${\rm ISM_B}$; Sari, Piran, \& Narayan 1998)\nocite{spn98}, and (ii) a
jet with $\theta_{\rm jet}<\Gamma_{t\sim 0.06\,{\rm d}}^{-1}$ (i.e.~a
jet break prior to the first observation at $t\approx 0.06$ days;
hereafter, ${\rm Jet_B}$).  The subscript B indicates that $\nu_c$ is
blueward of the optical bands initially.  In both models we use
Eqn.~\ref{eqn:model2} for the $R$-band lightcurve, with $t_b$ defined
as the time at which $\nu_c=\nu_R$, and $\alpha_2\equiv\alpha_1-1/4$.

We find that in the ${\rm ISM_B}$ model $t_c\approx 0.4$ days, while 
in the ${\rm Jet_B}$ model $t_c\approx 0.65$ days.  Moreover, in both  
models the closure relations can only be satisfied by including a
contribution from dust extinction within the host galaxy, $A_V^{\rm
host}$.  We estimate the required extinction using the parametric
extinction curves of Cardelli, Clayton, \& Mathis (1989)\nocite{ccm89}
and Fitzpatrick \& Massa (1988)\nocite{fm88}, along with the
interpolation calculated by Reichart (2001)\nocite{rei01}.  Since the
redshift of \grb{} is not known we assume $z=0.3,\,1,\,3$, which spans
the range of typical redshifts for the long-duration GRBs.  The
inferred values of $A_V^{\rm host}$ are summarized in
Tab.~\ref{tab:models}, and range from 0.2 to 0.9 mag.

\subsection{Jet Break}
\label{sec:jet}

An alternative explanation for the steepening is a jet expanding into:
(i) an ISM medium with $\nu_c$ blueward of the optical bands (J-${\rm
ISM_B}$), (ii) a Wind medium with $\nu_c$ blueward of the optical
bands (J-${\rm Wind_B}$), and (iii) an ISM or Wind medium with $\nu_c$
redward of the optical bands (J-${\rm ISM/Wind_R}$).  We note that the
J-${\rm ISM_B}$ model is different than the ${\rm ISM_B}$ model
(\S\ref{sec:cooling}) since previously it was defined such that the
jet break is later than the last observation.  In these models,
$t_b\equiv t_{\rm jet}$ is the time at which $\Gamma(t_{\rm
jet})\approx \theta_{\rm jet}^{-1}$.

From the closure values we note that the J-${\rm ISM/Wind_R}$ requires
no extinction within the host galaxy, while the J-${\rm ISM_B}$ and 
J-${\rm Wind_B}$ models require values of about $0.05$ to $0.3$ mag.

We find $t_{\rm jet}\sim 10-20$ days, corresponding to $\theta_{\rm
jet}\sim 10^\circ$.  Using the measured fluence (\S\ref{sec:opt}) we
estimate the beaming-corrected $\gamma$-ray energy, $E_\gamma\approx
5\times 10^{50}$ erg, assuming a circumburst density of 1 cm$^{-3}$
and $z=1$ ($E_\gamma$ is a weak function of $z$).  This value is in
good agreement with the distribution of $E_\gamma$ for long-duration
GRBs (Frail et al. 2001).

\section{Discussion and Conclusions} 
\label{sec:conc}

\begin{figure} 
\plotone{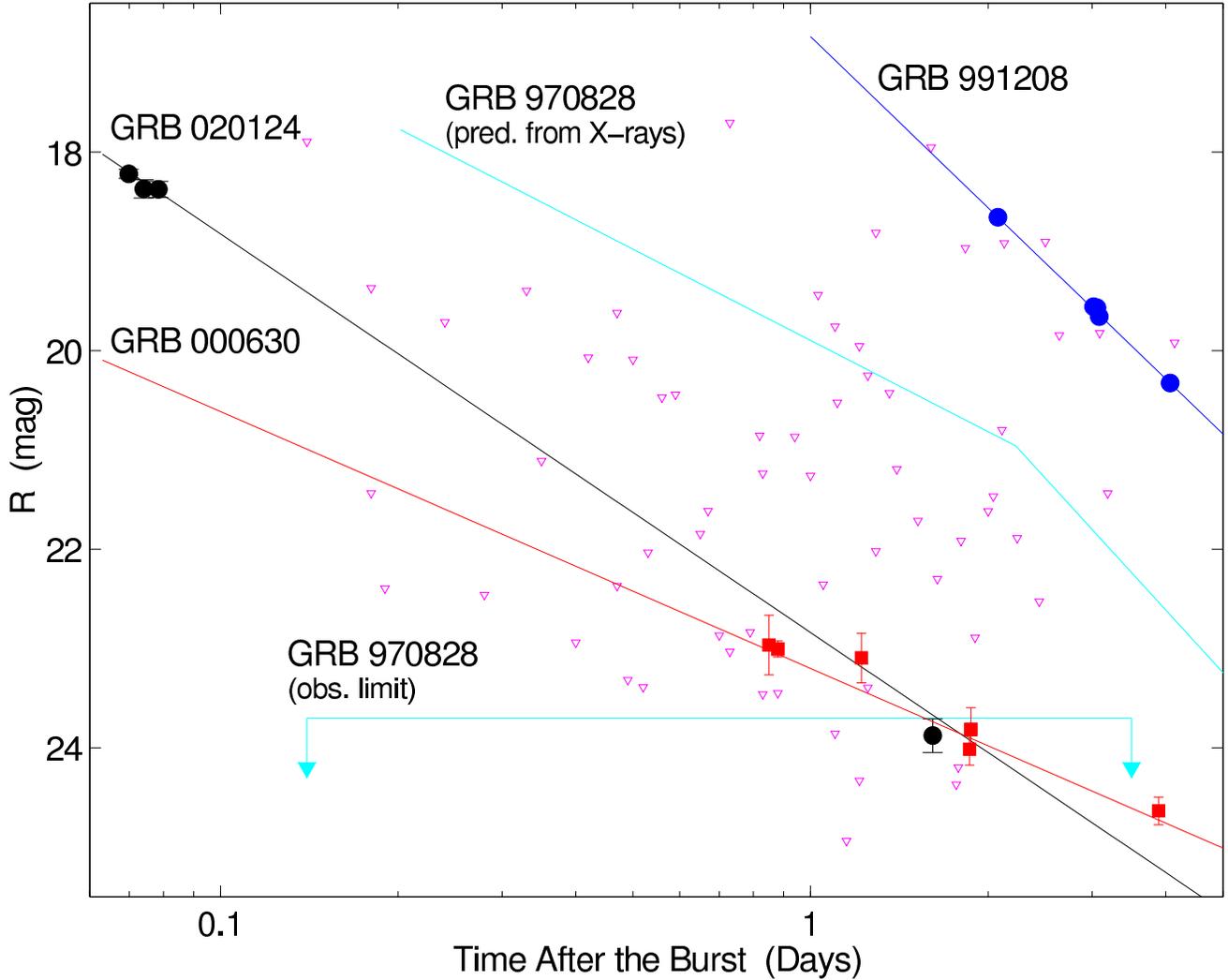}
\caption{$R$-band upper limits from searches of well-localized GRBs,
corrected for Galactic extinction.  The limits up to GRB\,000630 are
taken from Fynbo et al. (2001), while subsequent limits are from the
GRB Coordinates Network.  Also shown are the lightcurves of the
\grb{}, GRB\,000630, the bright GRB\,991208 (Castro-Tirado et
al. 2001), and GRB\,970828 (the de-reddened lightcurve is based on the
radio and X-ray data; Djorgovski et al. 2001).  Only about $30\%$ of
the searches yielded limits that are fainter than the afterglow of
\grb{}.  A similar fraction was found by Fynbo et al. (2001) based on
the afterglow of GRB\,000630.  
\label{fig:rlims}}
\end{figure}

Regardless of the specific model for the afterglow emission, the main
conclusion of \S\ref{sec:model} is that the optical afterglow of
\grb{} suffered little or no dust extinction.  Still, this afterglow
would have been missed by typical searches undertaken even as early as
12 hours after the GRB event.  As shown in Fig.~\ref{fig:rlims}, about
$70\%$ of the searches conducted to date would have failed to detect
an optical afterglow like that of \grb{}.

\begin{figure} 
\plotone{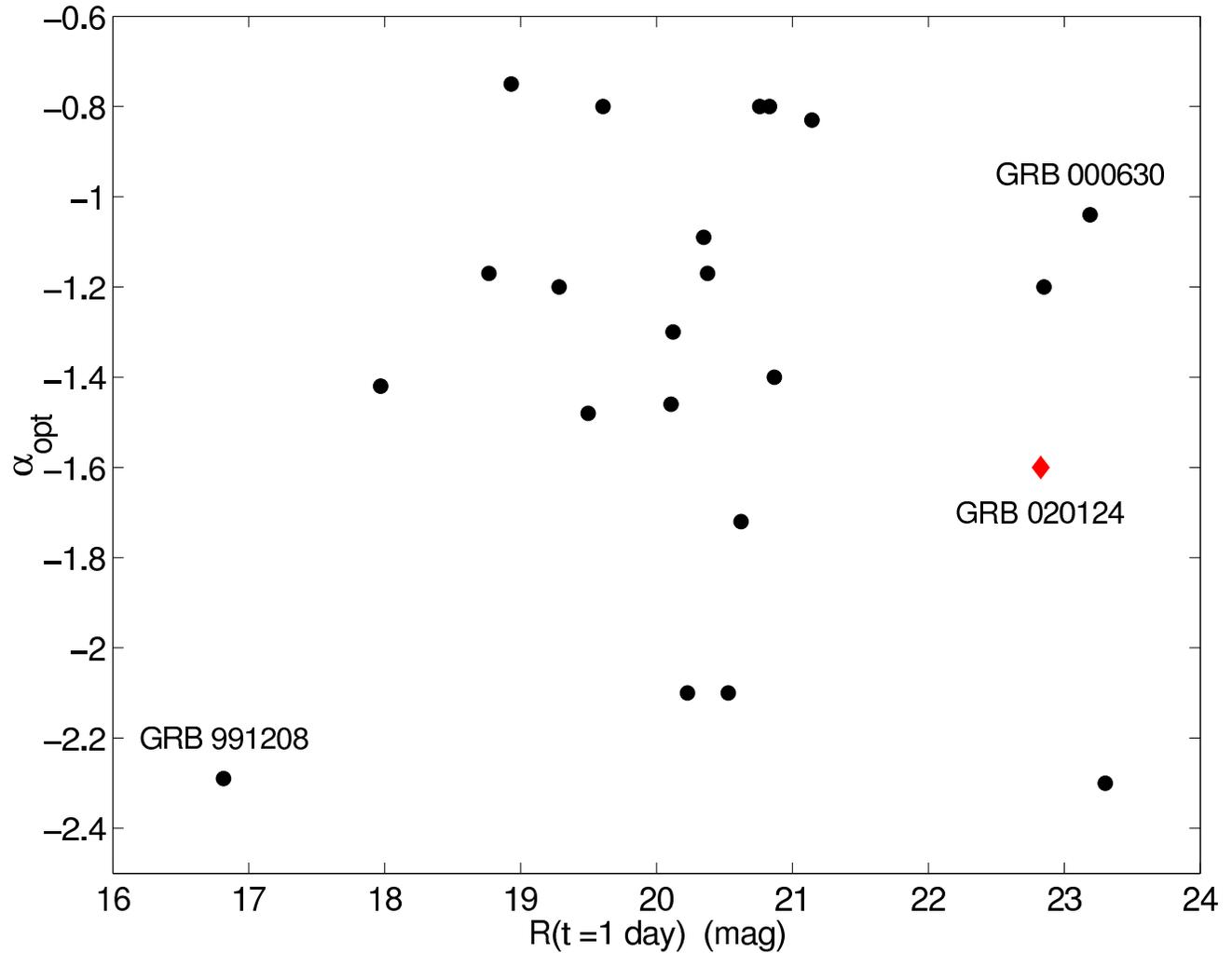}
\caption{Temporal decay index, $\alpha_{\rm opt}$ ($F_\nu\propto
t^\alpha$), plotted against the $R$-band magnitude at $t=1$ day for
several optical afterglows.  We chose a fiducial time of 1 day since,
with the exception of GRB\,010222, all the observations are before the
jet break.  While the majority of optical afterglows cluster around
$R(t=1\,{\rm d})\sim 20$ mag, \grb{} is one of the four faintest
afterglows detected to date, and one of the six most rapid faders.
\label{fig:alpha_f0}}
\end{figure}

This is simply because the afterglow of \grb{} was faint and exhibited
relatively rapid decay.  From Fig.~\ref{fig:alpha_f0} we note that
\grb{} is one of the faintest afterglows detected to date (normalized
to $t=1$ day), and while it is not an excessively rapid fader, it is
in the top $30\%$ in this category.  

Thus, the afterglow of \grb{}, along with that of GRB\,000630 (Fynbo
et al. 2001; Fig.~\ref{fig:alpha_f0}), indicates that there is a wide
diversity in the brightness and decay rates of optical afterglows.  In
fact, the brightness distribution spans a factor of about 400, while
the decay index varies by more than a factor of three.  Coupled with
the low dust extinction in the afterglow of \grb{}, this indicates
that some dark bursts may simply be dim, and not dust obscured.  

Given this wide diversity in the brightness of optical afterglows,
it is important to establish directly that an afterglow is dust
obscured.  This has only been done in a few cases (\S\ref{sec:intro}).
Therefore, while {\it statistical} analyses (e.g.~Reichart \& Yost
2001) point to extinction as the underlying reason for some fraction
of dark bursts, and may even account for an afterglow like that of
\grb{}, it is clear that observationally the issue of dark bursts is
not settled, and the observational biases have not been traced fully.

Since progress in our understanding of dark bursts will benefit from
observations, we need consistent, rapid follow-up of a large number of
bursts to constrain the underlying distribution, as well as
complementary techniques which can directly measure material along the
line of sight.  This includes X-ray observations which allow us to
measure the column density to the burst (Galama \& Wijers
2001)\nocite{gw01}, and thus infer the type of environment, and
potential extinction level.  Along the same line, radio observations
allow us to infer the synchrotron self-absorption frequency, which is
sensitive to the ambient density (e.g.~Sari \& Esin
2001)\nocite{se01}; the detection of radio emission, as in the case of
\grb{}, implies a density $n\lesssim 10^2$ cm$^{-3}$.  Finally, prompt
optical observations, as we have carried out in this case, may uncover
a larger fraction of the dim optical afterglows, and provide a better
constraint on the fraction of truly obscured bursts.

\acknowledgements 
J.~S.~B.~is a Fannie and John Hertz Foundation
Fellow. F.~A.~H.~acknowledges support from a Presidential Early Career
award. S.~R.~K.~and S.~G.~D.~thank the NSF for support.  R.~S.~is
grateful for support from a NASA ATP grant.  R.~S.~and
T.~J.~G.~acknowledge support from the Sherman Fairchild Foundation.
J.~C.~W.~acknowledges support from NASA grant NAG59302.  K.~H. is
grateful for Ulysses support under JPL contract 958056 and for IPN
support under NASA grants FDNAG 5-11451 and NAG 5-17100.  Support for
Proposal HST-GO-09180.01-A was provided by NASA through a grant from
the Space Telescope Science Institute, which is operated by the
Association of Universities for Research in Astronomy, Inc., under
NASA contract NAS5-26555.

\clearpage
\begin{deluxetable}{lccl}
\tabcolsep0.1in\footnotesize
\tablewidth{\hsize}
\tablecaption{Ground-Based Optical Observations of \grb{}\label{tab:opt}} 
\tablehead {
\colhead {UT}        &
\colhead {Telescope} &
\colhead {Band}      &
\colhead {Magnitude}
}
\startdata
Jan 24.51204 & MSO\,50  &  $R_M$  &    $17.918\pm 0.041$ \\
Jan 24.51204 & MSO\,50  &  $B_M$  &    $18.628\pm 0.057$ \\
Jan 24.51516 & SSO\,40  &  $R$    &    $18.219\pm 0.046$ \\
Jan 24.51655 & MSO\,50  &  $R_M$  &    $17.984\pm 0.044$ \\
Jan 24.51655 & MSO\,50  &  $B_M$  &    $18.727\pm 0.063$ \\
Jan 24.51938 & SSO\,40  &  $R$    &    $18.371\pm 0.091$ \\
Jan 24.52106 & MSO\,50  &  $R_M$  &    $18.111\pm 0.049$ \\
Jan 24.52106 & MSO\,50  &  $B_M$  &    $18.842\pm 0.069$ \\
Jan 24.52373 & SSO\,40  &  $R$    &    $18.376\pm 0.082$ \\
Jan 24.55791 & MSO\,50  &  $R_M$  &    $18.678\pm 0.048$ \\
Jan 24.55791 & MSO\,50  &  $B_M$  &    $19.661\pm 0.090$ \\
Jan 24.56243 & MSO\,50  &  $R_M$  &    $18.867\pm 0.036$ \\
Jan 24.56243 & MSO\,50  &  $B_M$  &    $19.584\pm 0.053$ \\
Jan 24.56696 & MSO\,50  &  $R_M$  &    $18.843\pm 0.039$ \\
Jan 24.56696 & MSO\,50  &  $B_M$  &    $19.714\pm 0.050$ \\
Jan 26.34100 & P\,200   &  $r'$   &    $24.398\pm 0.228$ 
\enddata
\tablecomments{The columns are (left to right), (1) UT date of
each observation, (2) telescope (MSO\,50: Mt.~Stromlo Observatory
50-in; SSO\,40: Siding Spring Observatory 40-in; P\,200: Palomar
Observatory 200-in), (3) observing band, and (4) magnitudes and
uncertainties.  The observed magnitudes are not corrected for Galactic
extinction.} 
\end{deluxetable}

\clearpage
\begin{deluxetable}{lccccc}
\tabcolsep0.1in\footnotesize
\tablewidth{\hsize}
\tablecaption{HST/STIS Observations of \grb{}\label{tab:opt2}} 
\tablehead {
\colhead {Epoch}        &
\colhead {Band}      &
\colhead {Exp.~Time} &
\colhead {Flux} &
\colhead {S/N} &
\colhead {Magnitude} \\
\colhead {(UT)} &
\colhead {} &
\colhead {(ksec)} &
\colhead {(e$^-$ s$^{-1}$)} &
\colhead {} &
\colhead {} 
}
\startdata
Feb 11.09 & $50\,$CCD/Clear & 10.0 & $0.0814\pm 0.0169$ & 4.82 & $R=28.68^{+0.25}_{-0.20}$ \\
Feb 18.30 & $50\,$CCD/Clear & 7.4  & $0.0443\pm 0.0189$ & 2.34 & $R=29.35^{+0.60}_{-0.39}$ \\
Feb 25.71 & $50\,$CCD/Clear & 7.5  & $0.0362\pm 0.0183$ & 1.98 & $R=29.56^{+0.76}_{-0.44}$ \\\hline
Feb 18.30+25.71 & $50\,$CCD/Clear & 14.9 & $0.0398\pm 0.0137$ & 2.91 & $R=29.46^{+0.46}_{-0.32}$
\enddata
\tablecomments{The columns are (left to right), (1) UT date of
each observation, (2) STIS CCD mode, (3) exposure time, (4) flux and
uncertainty, (5) significance, and (6) $R$ magnitude and uncertainty.
The total number of counts was converted to the $R$-band assuming the
observed color of the OT, $f_\lambda \propto \lambda^{-0.4}$
(\S\ref{sec:hst}).  For epochs 2 and 3, the $3\sigma$ upper limits
are: $R=29.09$ and $R=29.13$ mag, respectively.  The observed
magnitudes are not corrected for Galactic extinction.} 
\end{deluxetable}

\clearpage
\begin{deluxetable}{llc}
\tabcolsep0.1in\footnotesize
\tablewidth{\hsize}
\tablecaption{VLA Radio Observations of \grb{} \label{tab:rad}}
\tablehead {
\colhead {Epoch}      &
\colhead {$\nu_0$} &
\colhead {Flux Density} \\
\colhead {(UT)}      &
\colhead {(GHz)} &
\colhead {($\mu$Jy)}
}
\startdata
Jan 26.22 & 8.46 & $84\pm 30$ \\
Jan 26.25 & 22.5 & $-60\pm 100$ \\
Jan 27.22 & 8.46 & $45\pm 25$ \\
Feb 1.40  & 8.46 & $49\pm 17$ \\\hline
Jan 26.22-Feb 1.40 & 8.46 & $48\pm 13$
\enddata
\tablecomments{The columns are (left to right), (1) UT date of
each observation, (2) observing frequency, and (3) flux density at the
position of the radio transient with the rms noise calculated from
each image.  The last row gives the flux density at 8.46 GHz from the
co-added map.} 
\end{deluxetable}

\clearpage
\begin{deluxetable}{lccccccc}
%\rotate
\tabcolsep0.1in\footnotesize
\tablewidth{\hsize}
\tablecaption{Afterglow Models \label{tab:models}}
\tablehead {
\colhead {Model}    &
\colhead {$\alpha_1$} &
\colhead {$\alpha_2$} &
\colhead {$\beta$}  &
\colhead {($b,c$)}  &
\colhead {Closure}  &
\colhead {$p$}      &
\colhead {$A_V^{\rm host}$ (mag)}
}
\startdata
${\rm ISM_B}$      & $-\frac{3(p-1)}{4}$ & $-\frac{3p}{4}+\frac{1}{2}$
& $-\frac{p-1}{2}$ & ($-3/2,0$)          & $0.52\pm 0.28$ & $3.17\pm
0.05$   & ($0.35,\,0.18,\,0.10$) \\
${\rm Jet_B}$      & $-p$                & $-p$ & $-\frac{p-1}{2}$ &
($-2,1$)           & $2.23\pm 0.36$      & $1.63\pm 0.04$ &
($0.89,\,0.50,\,0.22$) \\ \hline
J-${\rm ISM_B}$      & $-\frac{3(p-1)}{4}$ & $-p$ & $-\frac{p-1}{2}$ 
& ($-3/2,0$)       & $0.52\pm 0.28$      & $3.17\pm 0.05$   
& ($0.30,\,0.10,\,0.05$) \\
J-${\rm Wind_B}$     & $-\frac{3p-1}{4}$   & $-p$  & $-\frac{p-1}{2}$ &
($-3/2,1/2$)      & $1.02\pm 0.28$      & $2.51\pm 0.05$ &
($0.30,\,0.16,\,0.08$) \\
J-${\rm ISM/Wind_R}$ & $-\frac{3p-2}{4}$   & $-p$  & $-\frac{p}{2}$   &
($-3/2,-1/2$)       & $0.02\pm 0.28$      & $2.84\pm 0.05$ 
& \nodata
\enddata
\tablecomments{The columns are (left to right), (1) Afterglow model
(ISM: $r^0$ circumburst medium; Wind: $r^{-2}$ circumburst medium;
Jet: collimated eject with opening angle $\theta_{\rm jet}$; a 
subscript $B$ indicates $\nu_c<\nu_{\rm opt}$, and a subscript $R$
indicates $\nu_c>\nu_{\rm opt}$), (2) $\alpha_1$ as a function of $p$,
(3) $\alpha_2$ as a function of $p$, (4) $\beta$ as a function of $p$,
(5) closure relations ($\alpha+b\beta+c=0$), (6) resulting closure
values from the observed values of $\alpha_1$ and $\beta$, (7)
inferred value of $p$ from the measured value of $\alpha_1$, and (8)
the required extinction in the frame of the host galaxy for closure
values of zero ($z=0.3,\,1,\,3$); typical uncertainties are $\pm 0.05$
mag.  The top two models apply to the case when the observed
steepening in the lightcurves is due to the passage of $\nu_c$ through
the $R$-band, while the bottom three apply to the case when the
steepening is due to a jet.}  
\end{deluxetable}

\clearpage

\clearpage

\clearpage

\clearpage

\clearpage

\end{document}